\documentclass[twocolumn,shortnote]{jpsj2}
\usepackage{graphicx, color}
\def\runauthor{Online-Journal Subcommittee}

\title{
Bulk superconductivity in Bi$_4$O$_4$S$_3$ revealed by specific heat measurement
}

\author{%
Hiroshi Takatsu$^{1,}$\thanks{E-mail address: takatsu@tmu.ac.jp},
Yoshikazu Mizuguchi$^2$, Hiroki Izawa$^2$, Osuke Miura$^2$, and  Hiroaki Kadowaki$^1$
}
\inst{%
$^1$Department of Physics, Tokyo Metropolitan University, Hachioji-shi, Tokyo 192-0397, Japan

$^2$Department of Electrical and Electronic Engineering, Tokyo Metropolitan University, Hachioji-shi, Tokyo 192-0397, Japan
}
\recdate{\today}

\kword{%
Bi$_4$O$_4$S$_3$, Specific heat
}

\begin{document}
\maketitle

Since the recent discovery of superconductivity in Bi$_4$O$_4$S$_3$, 
the BiS$_2$-based superconducting family has attracted many 
researchers~\cite{Y.Mizuguchi-condmat-1,Y.Mizuguchi-condmat-2,S.Demura-condmat-1,S.Li-condmat-1,S.G.Tan-condmat-1,Awana-condmat-1,Singh-condmat-1,H.Kotegawa-condmat-1,H.Usui-condmat-1,T.Zhou-condmat-1,X.Wan-condmat-1}
because of some analogies to cuprates and Fe-based superconductors~\cite{J.G.Bednorz1986,IshidaJPSJ2009,MizuguchiJPSJ2010}.
The parent phase of the Bi$_4$O$_4$S$_3$ superconductor is Bi$_6$O$_8$S$_5$ with 
a crystal structure composed of a stacking of Bi$_2$S$_4$ layers 
(two BiS$_2$ layers)
and Bi$_4$O$_4$(SO$_4$) spacer layers~\cite{Y.Mizuguchi-condmat-1}. 
Band calculations~\cite{Y.Mizuguchi-condmat-1,H.Usui-condmat-1} indicate that Bi$_6$O$_8$S$_5$ is an insulator with Bi$^{3+}$.
Superconducting Bi$_4$O$_4$S$_3$ is expected to possess 50\% defects at a SO$_4$ site,
which generates electron carriers within the BiS$_2$ layers. 
In fact, superconductivity of Bi$_4$O$_4$S$_3$ is induced by electron doping into 
the BiS$_2$ layers via the defects of the SO$_4$ ions at the interlayer site~\cite{Y.Mizuguchi-condmat-1}. 
To date, LaOBiS$_2$ and NdOBiS$_2$, having analogous BiS$_2$ layers, 
have been found to show superconductivity by electron doping as well~\cite{Y.Mizuguchi-condmat-2,S.Demura-condmat-1}.
Although the nature of superconductivity in the BiS$_2$ family has not been clarified so far, 
the relatively high transition temperature ($T_\mathrm{c}$), for example, 
10.6~K in La(O,F)BiS$_2$, attracts researchers to exploring new BiS$_2$-based superconductors with a higher $T_\mathrm{c}$. 
To elucidate the superconductivity mechanism or design new BiS$_2$-based superconductors, 
it is important to clarify whether the observed superconductivity is bulk or occurred in surface. 
Here we show the evidence of
bulk superconductivity in Bi$_4$O$_4$S$_3$ revealed by the specific heat measurements.

%
The polycrystalline Bi$_4$O$_4$S$_3$ sample was prepared by the conventional solid-state reaction method as described in Ref.~1. 
The sample quality was confirmed by the x-ray diffraction and dc susceptibility measurements. 
It was comparable to the result reported in Ref.~1.
The $T_\mathrm{c}$ was estimated to be $\sim4.7$~K.
Specific heat was measured with a thermal relaxation method with a commercial calorimeter (PPMS, Quantum Design)
down to 2~K.

\begin{figure}[t]
\begin{center}
 \includegraphics[width=0.45\textwidth]{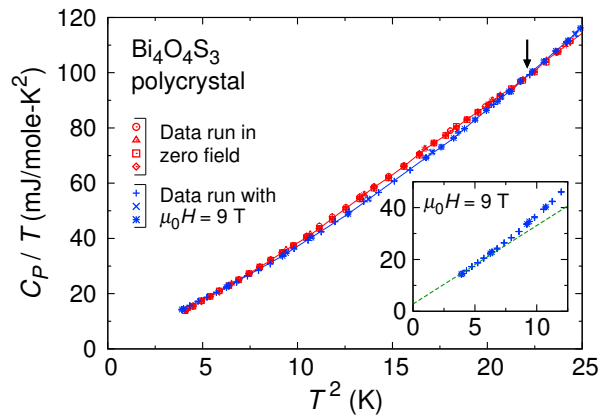}
\caption{
Temperature dependence of total specific heat $C_{P}$ of 
a polycrystalline sample of Bi$_4$O$_4$S$_3$.
The inset shows the same $C_P/T$ versus $T^2$ plot in the low-$T$ region at 9~T (normal state).
}
\label{fig.1}
\end{center}
\end{figure}

\begin{figure}[t]
\begin{center}
 \includegraphics[width=0.45\textwidth]{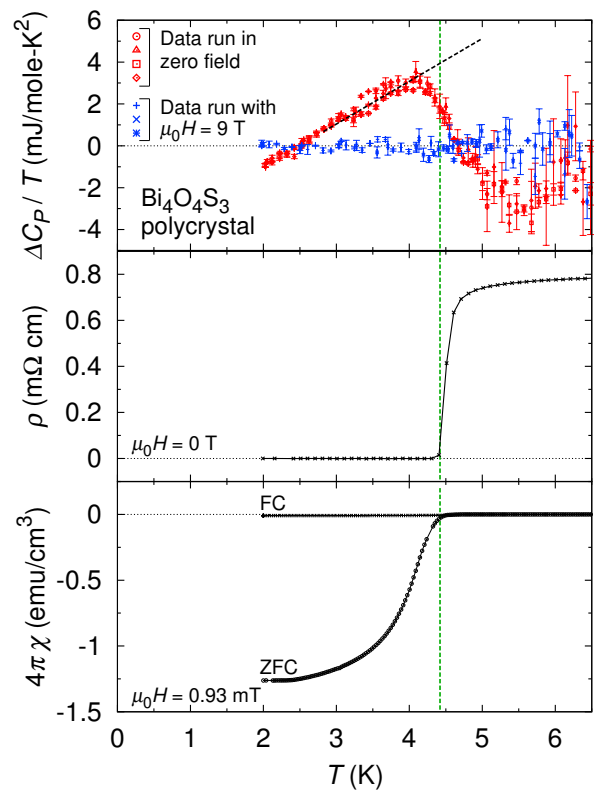}
\caption{
Temperature dependence of 
(a) the difference in the electronic specific heat 
between the superconducting state and the normal-conducting state, 
$\varDelta C_P = C_P(0\mathrm{T})-C_P(9\mathrm{T})$,
(b) resistivity $\rho$, and 
(c) dc magnetic susceptibility $\chi$.
The lines in (a) are entropy-conserving constructions in order to estimate the intrinsic jump height 
of $\varDelta{C_{P}}$.
The transition temperature estimated by $C_P$ is consistent with that from $\rho$ and $\chi$.
}
\label{fig.2}
\end{center}
\end{figure}

Figure~\ref{fig.1} shows the temperature dependence of the total specific heat $C_{P}$ 
in the superconducting state ($\mu_0H = 0$~T) and in the normal-conducting state 
($\mu_0H = 9~\mathrm{T} > \mu_0H_c$).
The $C_{P}$ anomaly associated with the superconducting transition 
is observed, as indicated by the arrow ($T \simeq 4.7$~K). 
This result ensures that Bi$_4$O$_4$S$_3$ is a bulk superconductor. 
The tiny jump is attributed to the small electronic-specific-heat coefficient $\gamma$.
Assuming the BCS weak coupling approximation,
$\varDelta{C_{P}}/\gamma{T_\mathrm{c}} = 1.43$ and 100\% superconducting volume,
the $\gamma$ value was yielded to be about 2.8~mJ/(f.u. mol K$^2$).
The inset of Fig.~\ref{fig.1} presents the fitting result by a conventional relation, $C_P/T = \gamma + \beta T^2$,
with the estimated $\gamma$ and the coefficient of the phononic contribution $\beta$. 
Although lower-$T$ data need to evaluate the certain $\gamma$ value,
the data is fairly fitted by the relation with 
$\beta = 3.0$~mJ/(f.u. mol K$^4$) yielding the Debye temperature 
$\varTheta_\mathrm{D} = 192$~K.
We expect that the small $\gamma$ and low carrier\cite{S.Li-condmat-1} are essential for 
the superconducting mechanism of the BiS$_2$-based superconductor.
%

%
From the intrinsic jump of $C_{P}$,
we define the transition temperature $T_{\mathrm{c}} = 4.4$~K in zero field,
which agrees well with the temperature of the zero-resistivity and the starting temperature 
of the bifurcation between $\chi_{\mathrm{FC}}$ and $\chi_{\mathrm{ZFC}}$.
In Fig.~\ref{fig.2},
comparisons between those results are presented.
For the $C_{P}$ data,
the difference in the electronic specific heat 
between the superconducting and normal-conducting states is estimated by a relation of
$\varDelta C_P = C_P(0\mathrm{T})-C_P(9\mathrm{T})$,
since the phononic contribution in $C_{P}$ is usually $H$ independent.
We have fitted the 9~T data to a polynomial, and used it in calculating 
$\varDelta{C_{P}}$. 
We obtained the intrinsic jump height at $T_{\mathrm{c}}$ as $\varDelta{C_{P}}/T_{\mathrm{c}} = 4.0$~mJ/(f.u. mol K$^2$).
We note that there is a small discrepancy between $C_{P}$ data of 
the superconducting and normal-conducting states above $T_{\mathrm{c}}$.
However, this is only 2\% of the total specific heat.
Therefore, we think that this is not intrinsic in the nature of Bi$_4$O$_4$S$_3$,
instead it is  
attributable to a small error of measurements or of the subtraction of the normal state contribution,
because of  the small specific heat jump.

In conclusion,
our specific heat experiments on a polycrystalline sample of the BiS$_2$-based superconductor Bi$_4$O$_4$S$_3$
demonstrate that the superconductivity of it is bulk in nature.
The $T_{\mathrm{c}}$ is estimated by the specific heat measurements to be $T_{\mathrm{c}} = 4.4$~K,
consistent with that of the resistivity and dc susceptibility.
In order to further discuss the superconductivity mechanism of Bi$_4$O$_4$S$_3$,
lower temperature specific heat experiments below 2~K is essential. 

\section*{Acknowledgment}
We acknowledge Yusuke Nakai and Yoshihiko Takano for fruitful discussions and experimental supports.
This work was partly supported by Grant-in-Aid for Research Activity Startup (23860042) from the Ministry of Education.

\bibliography{reference}
\end{document}